\documentclass[prb]{revtex4}
\usepackage{times}
\usepackage{wrapfig}
\usepackage{bm}
\usepackage{amsfonts}
\usepackage{amssymb}
\usepackage{psfrag}
\usepackage{graphicx}
\usepackage{amsmath}

\begin{document}

\title{Effective Free Energy of Ginzburg-Landau Model}

\author{Adriaan M.\ J.\ Schakel} 

\affiliation{National Chiao Tung University, Department of Electrophysics,
Hsinchu, 30050, Taiwan, R.O.C.}

\begin{abstract}
It is argued that the presence of a nonanalytic term in the effective
potential of the Ginzburg-Landau model is immaterial as far as the order
of the superconductor-normal phase transition is concerned.  To achieve
agreement with the renormalization group, the effective potential has to
be extended to include derivative terms, which provide the theory with a
low momentum scale which can be varied to probe the (possible) fixed
point. 
\end{abstract}

\maketitle

\section{Prelude}
This paper is dedicated to Professor Kleinert on the occasion of his
60th birthday with special thanks for the exciting years I had the good
fortune to be his {\it Wissenschaftlicher Assistent} at the Freie
Universit\"at Berlin.  It is an implementation of one of the many neat
ideas he shared with me over the years.  This particular one he told me
about four years ago in the late summer of 1996.

It is a commonly held belief that the order of an equilibrium phase
transition can be inferred from the form of the effective potential at
criticality.  The presence of, for example, a cubic term in the order
parameter is taken as signalling a discontinuous transition, while the
absence of such a term is taken as indicating a continuous transition.
At the mean-field level, where fluctuations in the order parameter are
ignored, this is certainly true\cite{Landau}.  However, the question
about the order of a transition is settled in the full theory, and
fluctuations may well change the mean-field result.

In a renormalization-group (RG) approach, a continuous phase transition
is associated with an infrared-stable fixed point in the space of
coupling parameters characterizing the theory.  Sufficiently close to
the transition, such a fixed point acts as an attractor to which the
couplings flow when one passes to larger length scales by integrating
out field fluctuations of smaller length scales.  When no
infrared-stable fixed point is detected in this process, the transition
is discontinuous.

It is important to note that a fixed point is probed by changing a
scale.  The effective potential evaluated at criticality in itself is
inadequate for this as it lacks a scale.  In this contribution we show
that a scale can be introduced by extending the effective potential to
include derivative terms.  The resulting effective free energy contains
the same information as that obtained in RG.  Along the way we are able
to implement Kleinert's idea of finding a fixed point without
calculating flow functions first.
\section{Cubic Term}
To be specific we consider the superconductor-normal phase transition
described by the O($n$) Ginzburg-Landau model, which has been one of
Kleinert's research topics for many years, and to which a large part of
the first volume of his textbook\cite{GFCM} {\it Gauge Fields in
Condensed Matter} is devoted.  The model is specified by the free energy
density (in the notation of statistical physics)
\begin{equation}   \label{GL}
{\cal E} = \left|(\partial_{\mu} - i e A_\mu)\bm{\phi}\right|^2 +
m^2 |\bm{\phi}|^2 + \lambda |\bm{\phi}|^4 + \frac{1}{4}
F_{\mu \nu}^2 + \frac{1}{2 \alpha} (\partial_{\mu} A_{\mu})^2 ,
\end{equation}	
with a complex order parameter $\bm{\phi}$ having an even number $n$ of
real field components:
\begin{equation} \label{phi}
\bm{\phi} = \frac{1}{\sqrt{2}} \left( \begin{array}{c} \phi_1 + i \phi_2 \\
\vdots \\ \phi_{n-1} + i \phi_n \end{array} \right).
\end{equation} 
A conventional superconductor with Cooper pairing in the s-channel
corresponds to $n=2$.  The parameters $e$ and $m$ are the electric
charge and mass of the field, while $\lambda$ is the coupling constant
characterizing the 4th-order interaction term.  We included a
gauge-fixing term with parameter $\alpha$, and $F_{\mu \nu} =
\partial_\mu A_\nu - \partial_\mu A_\nu$ is the (magnetic) field
strength.  For convenience we will work in the gauge $\partial_\mu A_\mu
=0$, which is implemented by taking the limit $\alpha \to 0$.  The mass
term depends on the temperature $T$, and changes sign at the critical
temperature $T_{\rm c}$.  In the Ginzburg-Landau model, $m^2 =
\xi_0^{-2}(T/T_{\rm c}-1)$, where $\xi_0$ is the length scale of
fluctuations in the amplitude of the order parameter.

\begin{figure}
\vspace{-.5cm}
\begin{center}
\includegraphics[width=6cm]{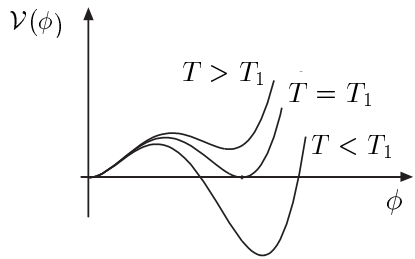}
\end{center}
\vspace{-0.2cm}
\caption{Sketch of the 1-loop effective potential.}
\label{fig:pot}
\end{figure}
As a first step to investigate the effect of fluctuations on the
mean-field picture, we consider, following Halperin, Lubensky, and
Ma\cite{HLM}, those in the vector field $A_\mu$.  Since (at least in the
normal phase) the corresponding mode is gapless, it can have an
important impact on the infrared behavior of the theory.  The functional
integral over $A_\mu$ is a simple Gaussian and gives as contribution to
the effective potential energy density in $d$ dimensions
\begin{equation} \label{Veff}
{\cal V}_{\rm eff}  = \frac{d-1}{2} \int \frac{d^d k}{(2 \pi)^d}
\ln(k^2 + 2 e^2 |\Phi|^2),
\end{equation}
assuming that the order parameter is a nonfluctuating background field
denoted by $\Phi$.  In deriving this, it is used that the combination
\begin{equation} \label{Pi}
P_{\mu \nu}(k) = \delta_{\mu \nu} - \frac{k_\mu k_\nu}{k^2},
\end{equation} 
which appears in intermediate steps, is a projection operator satisfying
$P^2 = P$, and that its trace gives the number of transverse components:
${\rm tr}\, P = d-1$.  The momentum integral is easily carried out using
dimensional regularization, with the result
\begin{equation} \label{Vfinal}
{\cal V}_{\rm eff} = \frac{1}{(2 \pi)^{d/2}} \frac{d-1}{d} \Gamma(1
-d/2) e^d |\Phi|^d,
\end{equation} 
where $\Gamma(x)$ is the Gamma function.  For $d=3$, this leads to the
cubic term in the effective potential mentioned in the
introduction\cite{HLM}.  The 1-loop contribution is to be added to the
mean-field potential ${\cal V}_0 = m |\Phi|^2 + \lambda |\Phi|^4$, and
is often taken as indicating a discontinuous phase transition at
a---what Kleinert\cite{KlPLB83} likes to call---precocious temperature
$T_1$ above the mean-field critical temperature $T_{\rm c}$,
\begin{equation} 
\frac{T_1}{T_{\rm c}} = 1 + \frac{\xi_0^2}{18 \pi^2} 
\frac{e^6}{\lambda},
\end{equation} 
where the mass term is still positive, and the order parameter jumps
from zero to the finite value $|\Phi|^2 = (1/18 \pi^2) e^6/\lambda^2$
(see Fig.\ \ref{fig:pot}).

Since this result is obtained in perturbation theory, both $e$ and
$\lambda$ are assumed to be small.  This still leaves the ratio of the
two, or the so-called Ginzburg-Landau parameter $\kappa_{\rm GL}^2 =
e^2/\lambda$ undetermined.  This parameter separates type-II
($\kappa_{\rm GL} > 1/\sqrt{2}$) superconductors, which have a Meissner
phase where an applied magnetic field can penetrate the sample in the
form of quantized flux tubes, and type-I ($\kappa_{\rm GL} <
1/\sqrt{2}$) superconductors, for which these flux tubes become
unstable.  Neglecting fluctuations in the order parameter, as is done
here, is valid in the type-I regime, where $\kappa_{\rm GL}$ is small.
(In the opposite limit, of deep type-II superconductors where
$\kappa_{\rm GL}$ is large, fluctuations in the vector field can be
neglected instead.)  This then led to the conclusion that type-I
superconductors undergo a discontinuous phase transition\cite{HLM}.  It
should, however, be noted that fluctuations in the order parameter
produce at criticality also a cubic term in the effective potential.
According to this argument, this would imply that also type-II
superconductors should undergo such a transition.

As the result discussed in this section is independent of the number of
field components, it should be valid for any $n$, including large
numbers.  This opens the possibility to check it using a $1/n$
expansion.

\section{$1/n$ Expansion}
The $1/n$ expansion can be applied when the number $n$ of field
components is large, so that its inverse provides the theory with a
small parameter.  Contributions are then ordered not according to the
number of loops, as in the loop expansion, but according to the number
of factors $1/n$.  The leading contribution in $1/n$ due to fluctuations
in the vector field is obtained by dressing its correlation function
with arbitrary many bubble insertions, and summing the entire set of
Feynman diagrams\cite{AppelHeinz}.  The resulting series is a simple
geometrical one, which leads to the following change in the correlation
function:
\begin{equation} 
\frac{P_{\mu \nu}(k)}{k^2} \to \frac{P_{\mu \nu}(k)}{k^2 - b_d
n e^2 \, k^{d-2} },
\end{equation}
where $b_d = c_d/(d-1)$, with $c_d$ the 1-loop integral
\begin{equation} 
c_d = \int \frac{d^d k}{(2 \pi)^d} \frac{1}{k^2 (k +q)^2}
\biggr|_{q^2=1} = \frac{\Gamma(2-d/2) \Gamma^2(d/2-1)}{(4 \pi)^{d/2}
\Gamma(d-2)}.
\end{equation}
Below $d=4$, the infrared behavior has become less singular as a result
of including the bubble insertions.  Instead of the effective potential
(\ref{Veff}), we now obtain
\begin{equation} \label{Vlargen}
{\cal V}_{\rm eff} = \frac{d-1}{2} \int \frac{d^d k}{(2 \pi)^d}
\ln(k^2 + b_d e^2 k^{d-2} + 2 e^2 |\Phi|^2).
\end{equation}
For large fields, $|\Phi| > (b_d e^{d -2} )^{1/(4-d)}$, the main
contribution to the integral comes from large momenta, where $k^2 + b_d
e^2 k^{d-4} \approx k^2$, so that in this limit we recover the result
(\ref{Vfinal}) obtained in the loop expansion: ${\cal V}_{\rm eff}
\propto |\Phi|^d$.  In the limit of small fields, on the other hand, the
main contribution comes from small momenta, where $k^2 + b_d e^2 k^{d-4}
\approx b_d e^2 k^{d-4}$.  We then find ${\cal V}_{\rm eff} \propto
|\Phi|^{2d}$ instead, implying that in $d=3$, the $1/n$ expansion no
longer gives a cubic term in the effective potential.  [This argument
captures only those terms in the Taylor expansion of the logarithm in
the effective potential (\ref{Vlargen}) which diverge in the infrared.
These are the important ones for our purposes as they can produce
nonanalytic behavior.  The first terms in the expansion may be infrared
finite, depending on the dimensionality $d$, and have to be treated
separately.  But they always lead to analytic terms and are therefore of
no concern to us here.]

The $1/n$ expansion calls into question the validity of the result
obtained in the loop expansion as it corresponds to large field values
and it is not clear whether this is still in the realm of perturbation
theory.

In the next section, we show that by extending the calculation of the
effective potential to that of the effective free energy, which includes
derivative terms, matters become consistent and in agreement with RG.

\section{Effective Free Energy}
When computing the effective free energy, not only fluctuations in the
vector field are to be considered, but also those in the order
parameter. To this end, we set
\begin{equation} 
\bm{\phi}(x) = \Phi(x) + \tilde{\bm{\phi}}(x),
\end{equation} 
with $\Phi(x)$ a nonfluctuating background field, and integrate out the
fluctuating field $\tilde{\bm{\phi}}$.  The main difference with the
previous calculation of the effective potential is that the background
field is no longer assumed to be constant, but can vary in space.  We
therefore also have to analyze the derivative terms in the effective
theory.

Because of gauge invariance, the results will depend solely
on the absolute value $|\Phi|$.  Without loss of generality we can
therefore assume that the background field $\Phi$ has only one nonzero
component, say the first one, $\Phi = (v, 0, \cdots, 0)/\sqrt{2}$.
For convenience we will work at criticality and set the mass parameter
$m$ to zero.

A straightforward calculation yields in the gauge $\partial_\mu A_\mu
=0$ the 1-loop effective free energy
\begin{eqnarray}  \label{seff}
F_{\rm eff}[v] &=& \frac{1}{2} {\rm Tr} \ln \left(1 + 3 \lambda
\frac{1}{p^2} v^2 \right) + \frac{n-2}{2} {\rm Tr} \ln \left(1 +
\lambda \frac{1}{p^2} v^2 \right)  \\ && + \frac{1}{2}
{\rm Tr} \ln \left[ \left( \begin{array}{cc} 1 & 0 \\ 0 & \delta_{\mu
\rho}
\end{array} \right) + \frac{1}{p^2} \left( \begin{array}{cc} 1 & 0 \\ 0 &
P_{\mu \nu}(p)
\end{array} \right) \left( \begin{array}{cc} \lambda v^2 & e
\partial_\nu v \\ e \partial_\rho v & e^2 v^2 \delta_{\nu \rho }
\end{array} \right) \right], \nonumber
\end{eqnarray} 
where we ignored an irrelevant constant, and the trace Tr denotes the
trace tr over discrete indices as well as the integration over momentum
and space.  More precisely,
\begin{equation} 
{\rm Tr} \ln[1 + K(x,p)] = {\rm tr} \int d^dx \, \int \frac{d^d k}{(2
\pi)^d} \, {\rm e}^{- i k \cdot x} \ln[1 + K(x,p)] \, {\rm e}^{i k \cdot
x},
\end{equation}  
with $p_\mu = -i \partial_\mu$ the derivative operating on everything to
its right.  Since the background field is space-dependent, the integrals
in Eq.\ (\ref{seff}) cannot be evaluated in closed form, but only in a
derivative expansion \cite{Fraser}.

The first step in this scheme is to expand the logarithm in a Taylor
series.  Each term of the series contains powers of the derivative
$p_\mu$, which in the second step are shifted to the left using the
identity
\begin{equation} \label{commu}
f(x) p_\mu g(x) = ( p_\mu - i \partial_\mu)  f(x) g(x), 
\end{equation} 
where $f(x)$ and $g(x)$ are arbitrary functions.  The symbol
$\partial_\mu$ in the last term denotes the derivative which, in
contrast to $p_\mu$, operates {\em only} on the next object to its
right.  The next step is to repeatedly integrate by parts until all the
$p_\mu$ derivatives operate to the left.  They then simply produce
factors of $k_\mu$ as only the function $\exp(-i k \cdot x)$ appears at
the left.  In shorthand, for an arbitrary function $h(k)$:
\begin{equation} 
\int \frac{d^d k}{(2 \pi)^d} {\rm e}^{- i k \cdot x} h(p) f(x) = \int
\frac{d^d k}{(2 \pi)^d} {\rm e}^{- i k \cdot x}
h(-\!\!\stackrel{\leftarrow}{p}) f(x) = \int \frac{d^d k}{(2
\pi)^d} {\rm e}^{- i k \cdot x} h(k) f(x),
\end{equation} 
ignoring total derivatives.  In this way, all occurrences of the
derivative operator $p_\mu$ are replaced with a mere integration
variable $k_\mu$.  The function $\exp(i k \cdot x)$ at the right can now
be moved to the left where it is annihilated by the function $\exp(-i k
\cdot x)$.  The momentum integration can then, in principle, be
performed and the effective free energy takes the form of a space integral
over a local density $F_{\rm eff} = \int d^dx \, {\cal E}_{\rm eff}$.

Applied to the formal expression (\ref{seff}), the derivative expansion
yields for the quadratic terms in the expansion of the logarithms there,
the following terms in the effective free energy density:
\begin{eqnarray} \label{L1}
{\cal E}_{\rm eff}(|\Phi|) = && - e^4 (d-1) \int \frac{d^d k}{(2 \pi)^d
} \frac{1}{k^2} \frac{1}{(k-i \partial)^2} \left|\partial_\mu \Phi
\right|^2 \nonumber \\ && -(n+8) \lambda^2 \int \frac{d^d k}{(2 \pi)^d }
\frac{1}{k^2} \frac{1}{(k-i \partial)^2} |\Phi|^2 |\Phi|^2 \nonumber \\
&& - e^2 \int \frac{d^d k}{(2 \pi)^d } \frac{1}{k^2} P_{\mu \nu}(k)
\frac{1}{(k- i \partial)^2} P_{\nu \mu}(k -i \partial) |\Phi|^2
|\Phi|^2.
\end{eqnarray}
Assuming that the order parameter carries a momentum $\kappa$, we can
replace all occurrences of the derivative $-i \partial_\mu$ with
$\kappa_\mu$.  

It is important to note the absence of infrared divergences here since
$\kappa_\mu$ acts as an infrared cutoff.  Without such a cutoff, as in
the effective potential (\ref{Veff}), the individual terms in the Taylor
expansion cannot be integrated because of infrared divergences, and the
entire series has to be summed, which can lead to nonanalytic
contributions.  From the perspective of RG, the presence of the momentum
scale $\kappa_\mu$ is crucial as it allows the study of the fixed point
by letting that scale approach zero.

With the tree contribution ${\cal E}_{\rm tree}$ added, we obtain after
carrying out the integrals over the loop momentum
\begin{equation}  \label{eval}
{\cal E}_{\rm tree} + {\cal E}_{\rm eff} = \left[ 1 - c \, (d-1)
\hat{e}^2\right] \left|\partial_{\mu}\Phi\right|^2 + \left\{
\hat{\lambda} - c \left[ (n+8) \hat{\lambda}^2 + d(d-1) \hat{e}^4
\right] \right\} \kappa^{4-d} |\Phi|^4,
\end{equation} 
where $\hat{\lambda}= \lambda \kappa^{d-4}$ and $\hat{e}^2 = e^2
\kappa^{d-4}$ are the rescaled dimensionless coupling constants.  As an
aside, by evaluating the integrals in fixed dimension $2 < d <4$, and
not in an $\epsilon$ expansion close to the upper critical dimension
$d=4$, we in effect implement Parisi's approach to critical
phenomena\cite{Parisi}.  The factor in front of the kinetic term in Eq.\
(\ref{eval}) amounts to a field renormalization.  It can be absorbed by
introducing the renormalized field $\Phi_{\rm r} = Z_{\phi}^{-1/2}
\Phi$, with $Z_{\phi}$ the field renormalization factor
\begin{equation}   \label{phiZ}
Z_\phi = 1 + c_d \, (d-1) \hat{e}^2 .
\end{equation} 
The effective free energy density then becomes
\begin{equation} 
{\cal E}_{\rm tree} + {\cal E}_{\rm eff} = \left|\partial_{\mu}
\Phi_{\rm r} \right|^2 + \lambda_{\rm r} |\Phi_{\rm r}|^4,
\end{equation}
with $\lambda_{\rm r}$ the renormalized coupling
\begin{equation} \label{lasol}
\frac{1}{\hat{\lambda}} = \frac{1}{\hat{\lambda}_{\rm r}} - c_d \left[
(n+8) -2 (d-1) \frac{\hat{e}_{\rm r}^2}{\hat{\lambda}_{\rm r}} +
\frac{1}{4} d(d-1) \frac{\hat{e}_{\rm r}^4}{\hat{\lambda}_{\rm r}^2}
\right].
\end{equation} 
It was Kleinert's idea to put this equation in this form.  The reason
being that the critical point is approached by letting the momentum
scale approach zero.  Since the original coupling constant $\lambda$ is
fixed, the left side tends to zero when $\kappa \to 0$.  The resulting
quadratic equation then determines the value of $\hat{\lambda}_{\rm r}$
at the critical point, provided we know the value of $\hat{e}_{\rm r}^2$
there.  This procedure is equivalent to finding the root of the flow
equation for $\hat{\lambda}_{\rm r}$ in conventional RG\cite{Habil}.

To obtain the value of $\hat{e}_{\rm r}^2$ at criticality, we rescale
the vector field $A_\mu \to A_\mu /e $ and consider it instead of the
order parameter to be the background field.  The 1-loop effective free
energy obtained after integrating out the scalar fields reads:
\begin{equation}      
F_{\rm eff}[A] = \frac{n}{2} {\rm Tr} \ln \left[1+ \frac{1}{p^2} (2
p_\mu A_\mu - A_\mu^2) \right],
\end{equation} 
where we again ignored an irrelevant constant and used the gauge
$\partial_\mu A_\mu=0$.  The second term in the expansion of the
logarithm yields the first nonzero contribution
\begin{equation} 
{\cal E}_{\rm eff}(A) = n \int \frac{d^d k}{(2 \pi)^d }
\frac{k_\mu(k_\nu - i \partial_\nu)}{k^2 (k -i \partial)^2} A_\mu A_\nu.
\end{equation} 
If we assume that the vector field carries the same momentum $\kappa_\mu$
as does the order parameter, we obtain after carrying out the momentum
integral for the sum of the tree and the 1-loop contribution
\begin{equation} 
{\cal E}_{\rm tree} + {\cal E}_{\rm eff} = - \frac{1}{2 e_{\rm r}^2}
\partial^2 A_\mu A_\mu,
\end{equation} 
with $e_{\rm r}^2$ the renormalized coupling constant,
\begin{equation} 
\frac{1}{\hat{e}_{\rm r}^2} = \frac{1}{\hat{e}^2} + \frac{n}{2}
\frac{c_d}{d-1}.
\end{equation} 
Keeping $e$ fixed and letting $\kappa \to 0$ to approach the critical
point, we see that the renormalized coupling tends to a constant value
\begin{equation} \label{esol}
\hat{e}_{\rm r}^{* 2} = \frac{2}{n} \frac{d-1}{c_d}.
\end{equation} 
When this value is substituted in Eq.\ (\ref{lasol}) with the left side
set to zero, the resulting quadratic equation in $\hat{\lambda}_{\rm r}$
has two real solutions
\begin{equation} \label{tri} 
\hat{\lambda}_{\rm r}^* = \frac{1}{2n(n+8)} \frac{1}{c_d} \left[n + 4
(d-1)^2 \pm \Delta_d \right],
\end{equation} 
provided that the determinant
\begin{equation} \label{deltad}
\Delta_d := \sqrt{n^2 - 4 (d-2) (d-1)^2 (d+1) n - 16 (d-1)^3 (d+1)}
\end{equation} 
is real.  For fixed dimensionality, this condition is satisfied only for
a sufficient number of field components $n$.  Specifically, the minimum
number in $d=2,3,4$ is $n_{\rm c} (2) = 4\sqrt{3} \approx 6.9$, $n_{\rm
c}(3) = 16(2 + \sqrt{6}) \approx 71.2$, and\cite{HLM} $n_{\rm c}(4) =
12(15 + 4 \sqrt{15}) \approx 365.9$.  The plus sign in Eq.\
(\ref{deltad}) corresponds to the infrared-stable fixed point, while the
minus sign corresponds to the tricritical point, where the continuous
phase transition changes to a discontinuous one.  For a fixed number of
field components $n$, this 1-loop result shows that continuous behavior
is favored when $d \to 2$, while in the opposite limit, $d \to 4$,
discontinuous behavior is favored.
\section{Conclusions}
For a conventional 3-dimensional superconductor corresponding to $n=2$,
the 1-loop result fails to give an infrared-stable fixed point.  It
should, however, be kept in mind that there are infinitely many loop
diagrams, and the values of the coupling constants at criticality are
not particular small to justify low-order perturbation theory.  For
example, $\hat{e}_{\rm r}^{* 2} = 32/n$ according to Eq.\
(\ref{esol}) with $d=3$, which only for large $n$ is small.  Different
approaches are therefore required to investigate the presence of a fixed
point.

One is the dual approach\cite{dual,GFCM,KKS}, which is a formulation in
terms of magnetic vortex loops.  Monte Carlo simulations of the dual
model led to the conclusion that 3-dimensional type-II superconductors
undergo a continuous phase transition belonging to the XY universality
class with an inverted temperature axis\cite{DaHa}.  Using this
approach, Kleinert\cite{tricritical} predicted the presence of a
tricritical point at a value of the Ginzburg-Landau parameter
$\kappa_{\rm tri} \approx 0.8/\sqrt{2}$.

Another approach within the framework of the Ginzburg-Landau model
itself was put forward in Ref.~\cite{HeTe}, where---in our
language---the vector field was assumed to carry not the same momentum
$\kappa_\mu$ as the order parameter, but $\kappa_\mu/x$ instead.  The
charged fixed points for $n=2$ and $d=3$ are then located at
$\hat{e}_{\rm r}^{* 2} = 16/x$ which becomes small for large $x$,
facilitating the existence of an infrared-stable fixed point.  The free
parameter $x$ was determined by matching $\kappa_{\rm tri}$ with
Kleinert's estimate or Monte Carlo simulations\cite{Bartholomew} which
gave $\kappa_{\rm tri} \approx 0.42/\sqrt{2}$.  Form Eqs.\ (\ref{esol})
and (\ref{tri}) one obtains with the parameter $x$ included\cite{CMNS}
\begin{equation} 
\kappa^2_{\rm tri} = (8 + x + \sqrt{x^2 + 16 x - 176})/40.
\end{equation}   
The resulting values for the critical exponents are consistent with the
expected (inverted) XY universality class.

Also resummation techniques applied to results \cite{2loop} obtained to
2nd-order in the loop expansion predict the existence of an
infrared-stable fixed point in the Ginzburg-Landau model\cite{FoHo}.

In conclusion, by extending the effective potential to include
derivative terms, we achieved agreement with RG as they provide the
theory with a low momentum scale which can be varied to probe the fixed
point.  The presence of a nonanalytic term in the effective potential at
criticality was argued to be immaterial as far as the order of the phase
transition is concerned.

\section*{Acknowledgments}
I wish to thank B. Rosenstein for the kind hospitality at the Department
of Electrophysics, National Chiao Tung University, Hsinchu, Taiwan, and
for helpful discussions.  This work was funded by the National Science
Council (NCS) of Taiwan.
\end{document}